\documentclass[12pt]{article}
\usepackage{amssymb}
\usepackage{graphicx}
\usepackage{graphics}
\usepackage{cite}
\usepackage{ulem}
\usepackage{amsmath}
\usepackage{mathtools}    

\def\d{\partial}
\def\l{\left(}                                    
\def\r{\right)}

\newcommand{\be}{\begin{equation}}
\newcommand{\ee}{\end{equation}}
\newcommand{\ba}{\begin{align}}
\newcommand{\ea}{\end{align}}
\newcommand{\bg}{\begin{gather}}
\newcommand{\eg}{\end{gather}}
\newcommand{\bseq}{\begin{subequations}}
\newcommand{\eseq}{\end{subequations}}

\renewcommand{\tanh}{\mathop{\rm th}\nolimits}

\begin{document}
\title{Scalaron the healer: removing the strong-coupling in the Higgs-
and Higgs-dilaton inflations}

\author{Dmitry Gorbunov$^{1,2}$, Anna Tokareva$^{1}$\\
\mbox{}$^{1}${\small Institute for Nuclear Research of Russian Academy of
  Sciences, 117312 Moscow,
  Russia}\\  
\mbox{}$^{2}${\small Moscow Institute of Physics and Technology, 
141700 Dolgoprudny, Russia}\\ 
}

\maketitle
 
\begin{abstract}
We show that introducing $R^2$-term makes the Higgs-inflation and
Higgs-dilaton inflation consistent models: the strong coupling energy
scales in scalar, gauge and gravity sectors all are lifted up to the
Planck scale.
\end{abstract}
 
{\bf 1.} The Higgs-inflation \cite{Bezrukov:2007ep} is one of the physically motivated
inflationary models perfectly consistent with present cosmological
observations \cite{Ade:2015lrj}. It introduces a minimal modification to the
Standard Model of particle physics (SM) --- additional coupling of the
Higgs field to gravity. However, the model suffers from the {\it strong
  coupling problem} \cite{Burgess:2009ea,Barbon:2009ya}:
presently (in the electroweak vacuum) the model
becomes strongly coupled and loses the perturbative unitarity well
below the usual Planck scale, where gravity goes out of control and
hence the whole theory.  While the relevance of this observation for
validity of the inflationary solution \cite{Bezrukov:2007ep}
 is questionable
\cite{Bezrukov:2010jz,Bezrukov:2011sz}, the reheating, originally
estimated \cite{Bezrukov:2008ut,GarciaBellido:2008ab} to be within
the perturbative region actually tends to happen earlier
\cite{Ema:2016dny,DeCross:2016cbs}, when the theory is in the strong
coupling regime. Therefore,  
  a healthy  modification of the original model
free of the strong coupling is desirable. The models suggested so
far do not fully address these issues (see Ref.\,\cite{Bezrukov:2013fka} for
detailed discussion) leaving the problem unsettled.
 
In this {\it Letter} we put forward an idea that one of the most
natural modification of the Higgs-inflation --- achieved by adding a
quadratic in scalar curvature $R$ term --- allows one to push the
strong coupling scales in {\it all the model sectors} up to the
gravity scale $M_P$. This modification is a perturbatively tractable
inflationary model providing with robust predictions of the
cosmological parameters consistent with observations
\cite{Ade:2015lrj}.
 
{\bf 2.} While such a modification is natural within the quantum
perturbative theory in a model with non-minimal coupling to gravity
\cite{Avramidi:1986mj}, its capability of addressing the strong-coupling
issue can be illustrated as follows. We start with the action of the
Higgs-inflation augmented with the squared scalar curvature term,   
\be
\label{action-1}
S_0=\int d^4x \sqrt{-g} \l-\frac{M_P^2+\xi h^2}{2}R+\frac{\beta}{4}
R^2+\frac{(\d_{\mu}h)^2}{2}-\frac{\lambda}{4}(h^2-v^2)^2\r.
\ee
At $\beta=0$ eq.\,\eqref{action-1} describes the model of Higgs-inflation
($h$ stands for the Higgs field in unitary gauge). It 
successfully explains the flatness and homogeneity of the Universe and
produces the scalar and tensor perturbations consistent with the
present cosmological observations \cite{Ade:2015lrj}, provided
non-minimal coupling $\sim 10^3$-$10^4\gg 1$
\cite{Bezrukov:2007ep,Bezrukov:2009db}. However, this large coupling
is known to spoil the perturbative unitarity of the model at energy
scale $M_P/\xi$ \cite{Burgess:2009ea}.
 
With $\beta \ne 0$ the action \,\eqref{action-1} is known to provide with an extra scalar degree of freedom (scalaron) in the gravitational sector \cite{Stelle:1977ry}. The mass of this particle is given by 
\be 
\label{mass}
m=\frac{M_P}{\sqrt{3\beta}}\,.
\ee
 
Introducing a Lagrange multiplier $L$ and auxiliary scalar ${\cal R}$
we obtain from \eqref{action-1},
\be
S=\int{d^4 x \sqrt{-g}\, \l \frac{(\d_{\mu}h)^2}{2}-\frac{\lambda}{4}(h^2-v^2)^2-\frac{M_P^2+\xi h^2}{2}{\cal R}+\frac{\beta}{4} {\cal R}^2 -L {\cal R} +L R\r}.        
\ee
Then the field ${\cal R}$ can be integrated out,
\be
\label{action-3}
S=\int{d^4 x \sqrt{-g}\, \l \frac{(\d_{\mu}h)^2}{2}-\frac{\lambda}{4}(h^2-v^2)^2+L R - \frac{1}{\beta}(L+\frac{1}{2}\xi h^2+\frac{1}{2}M_P^2)^2\r},
\ee
and the large coupling $\xi$ is moved to the potential term. Moreover, the actual coupling constant is $\xi^2/\beta$ instead of $\xi$. This constant can well be smaller than unity provided \footnote{Note that the value $\beta\sim \xi^2$ is a natural choice of this parameter since non-minimal coupling $\xi$ induces a loop correction to $\beta$ of this order \cite{Avramidi:1986mj}.}
\be
\label{perturbativity}
\beta \gtrsim \frac{\xi^2}{4\pi}\,.
\ee
Thus, we expect that the problem corresponding to the large value of $\xi$ can be solved in this way. Below we show that this is indeed the case.
 
{\bf 3.}
Let us perform the Weyl transformation to the Einstein frame
\[
g_{\mu\nu}\to \Omega^2\,g_{\mu\nu}\,, \;\;\;
\Omega^2\equiv\frac{2L}{M_P^2}\,,
\]
and replace $L$ with $\phi$ (dubbed scalaron) introduced as
\[
\phi\equiv M_P\,\sqrt{\frac{2}{3}}\,\log\Omega^2\,.
\]
In terms of $h$, $\phi$ and the rescaled metric, action \eqref{action-3}
reads \cite{Ema:2017rqn}
\be
\label{action-4}
S=\int d^4 x \sqrt{-g}\, \l -\frac{R}{12} + \frac{1}{2}e^{-2\phi}(\d h)^2+\frac{1}{2}(\d\phi)^2-\frac{1}{4}e^{-4\phi}\l\lambda h^4+\frac{1}{36 \beta}(e^{2\phi}-1-6\xi h^2)^2\r\r
\ee
Hereafter we use a convention $M_P=1/\sqrt{6}$. In order to make the Higgs field canonical it is convenient to use a different set of variables,
\be
h=e^{\Phi}\tanh{H},~\phi=e^{\Phi}/\cosh{H},
\ee
which leads to the lagrangian {\it in the scalar sector} of the theory,  
\be
\label{action-5}
L=\frac{1}{2}\cosh^2{H}(\d \Phi)^2+\frac{1}{2}(\d H)^2- \frac{1}{4}\l\lambda\sinh^4{H}+\frac{1}{36 \beta}(1-e^{-2\Phi}\cosh^2{H}-6\xi\sinh^2{H})^2\r.
\ee
In this variables, the Higgs field couples to gauge $W$-bosons (and
similarly to $Z$-bosons) as
\be
\label{GB}
L_{gauge}=\frac{g^2 h^2}{4}e^{-2\phi} W^+_{\mu}W^-_{\mu}=\frac{g^2}{4} \sinh^2{H}~ W^+_{\mu}W^-_{\mu}.
\ee
 
{\bf 4.}
Let's determine positions of the strong coupling scales in the model
\eqref{action-1} in the Einstein frame. There are three sectors to be examined.
 
{\it Gravity sector.} Action \eqref{action-4} implicitly demonstrates
that gravity becomes strongly coupled at the Planck scale, as in
General Relativity. Indeed, the gravitons come from the curvature term,
which is of the standard form (Hilbert--Einstein, recall our
convention $M_P^2=1/6$).
 
{\it Scalar sector.} Here again, the model action in the form
\eqref{action-4} is useful. The interaction between Higgs field $h$
and scalaron $\phi$ is originated from the kinetic term (the second term
of lagrangian \eqref{action-4}) and the potential. Making use of the
series in field $\phi$ one finds, that it never comes with the large
coupling ($\phi$ always comes as $\phi/(\sqrt{6}M_P)$), so the kinetic
term in \eqref{action-4} is healthy up to the Planck scale. Similar is
true for the potential term of lagrangian \eqref{action-4}, provided
the inequality \eqref{perturbativity}, even if $\xi\gg 1$. This behavior
was also found in Ref.\,\cite{Ema:2017rqn}. Therefore, we conclude that with
model parameters obeying \eqref{perturbativity} the scalar is free from
the strong coupling problem up to the Planck scale, where scalaron
becomes strongly coupled.
 
\begin{figure}[!htb]
\begin{center}
\includegraphics[width=0.9\textwidth]{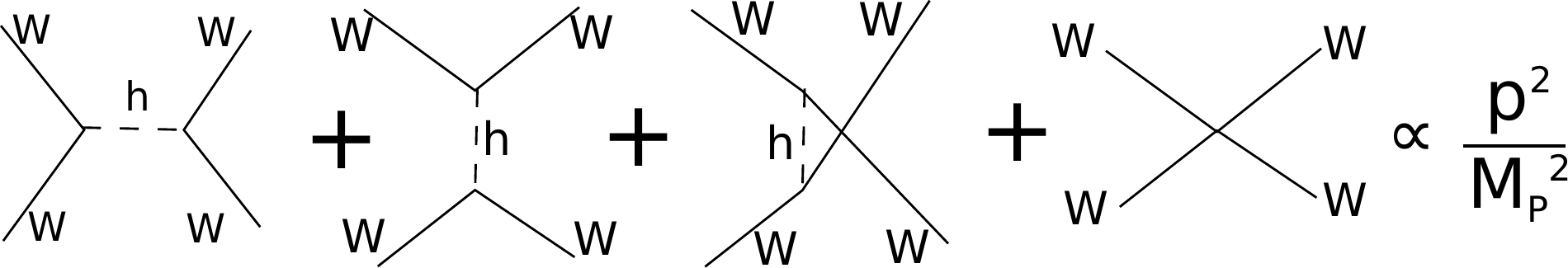}
\caption{Scattering of electroweak massive gauge bosons. 
\label{fig:can}}
\end{center}
\end{figure}

{\it Gauge sector.}
In the SM, the self-interaction of gauge bosons produces a
part of the $2\rightarrow 2$ scattering amplitude which grows with
the particle momenta above the electroweak scale, $\propto p^2/m_W^2$. However, this part coming from the scattering of
longitudinal modes is canceled by the vertices including the exchange
of the Higgs boson, see Fig.\,\ref{fig:can}.
 
If the Higgs sector is modified this compensation doesn't hold
anymore, that is the problem in the Higgs-inflation. The growing part of the amplitude can be written in the form,
\be
{\cal A}\sim \frac{g^2 p^2}{m_W^2}\l\frac{4}{g^2}\l\frac{d m_W(H)}{d H}\r^2-1\r\,.
\ee
In our model the canonically normalized Higgs field couples to the gauge bosons
via the term \eqref{GB} inducing the mass term of the form
$m_W=g\sinh{H}/2$ and hence 
\be
{\cal A}\propto \frac{p^2}{M_P^2}\,.
\ee
Thus, the unitarity cutoff scale for the scattering of the gauge
bosons is the Planck mass (here we restore the Planck mass, according to our convention $M_P=1/\sqrt{6}$).

In the fermionic sector of the Standard model, the scattering amplitude of two fermions to two gauge bosons will also grow linearly with the momentum (see \cite{Bezrukov:2012hx}).
\be
{\cal A}_f\sim y \,g\,\frac{p}{m_W}\l\frac{2}{g}\l\frac{d m_W(H)}{d H}\r-1\r\,\propto \frac{p}{M_P}\frac{\cosh H-1}{\sinh H}.
\ee
Thus, the corresponding unitarity cutoff scale in the fermionic sector is always higher than the Planck mass.
 
\begin{figure}[!htb]
\begin{minipage}[h]{0.4\linewidth}
\includegraphics[width=\textwidth]{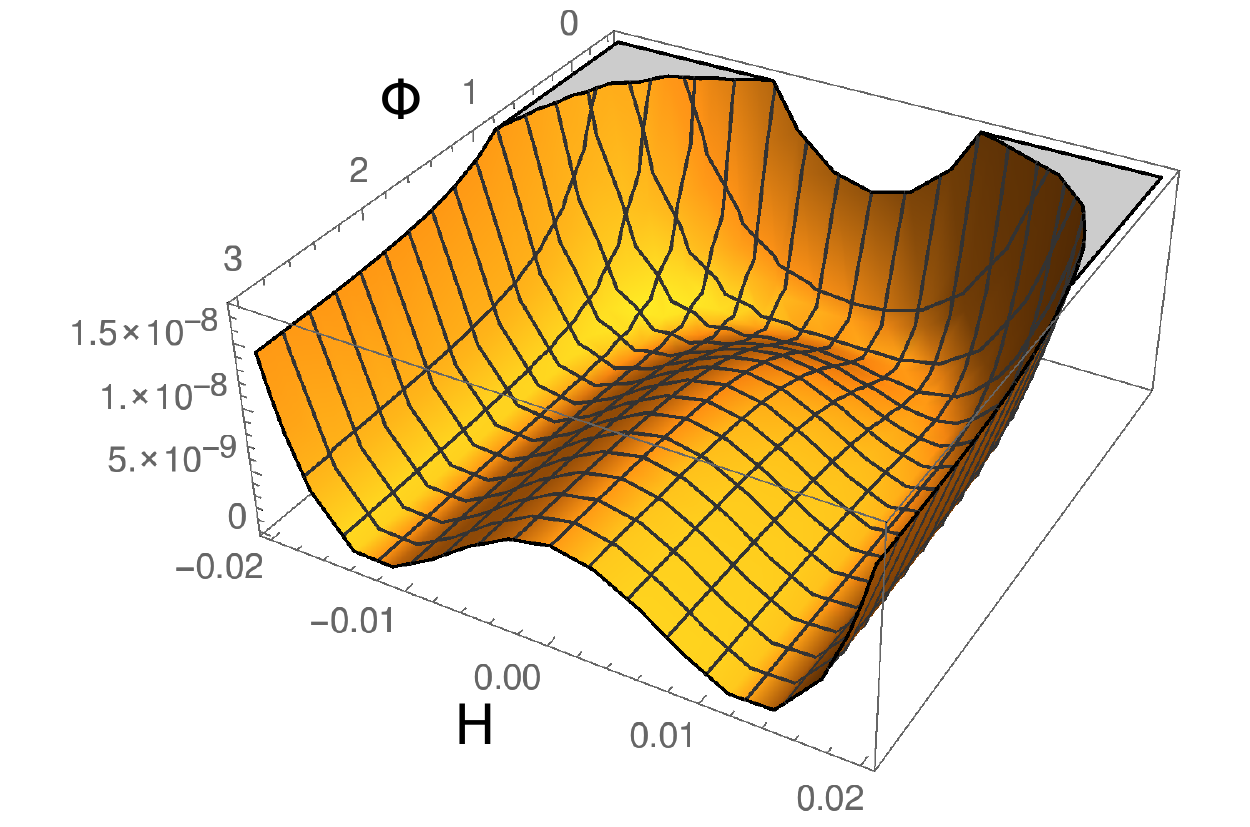}

\end{minipage}
\begin{minipage}[h]{0.6\linewidth}
\includegraphics[width=\textwidth]{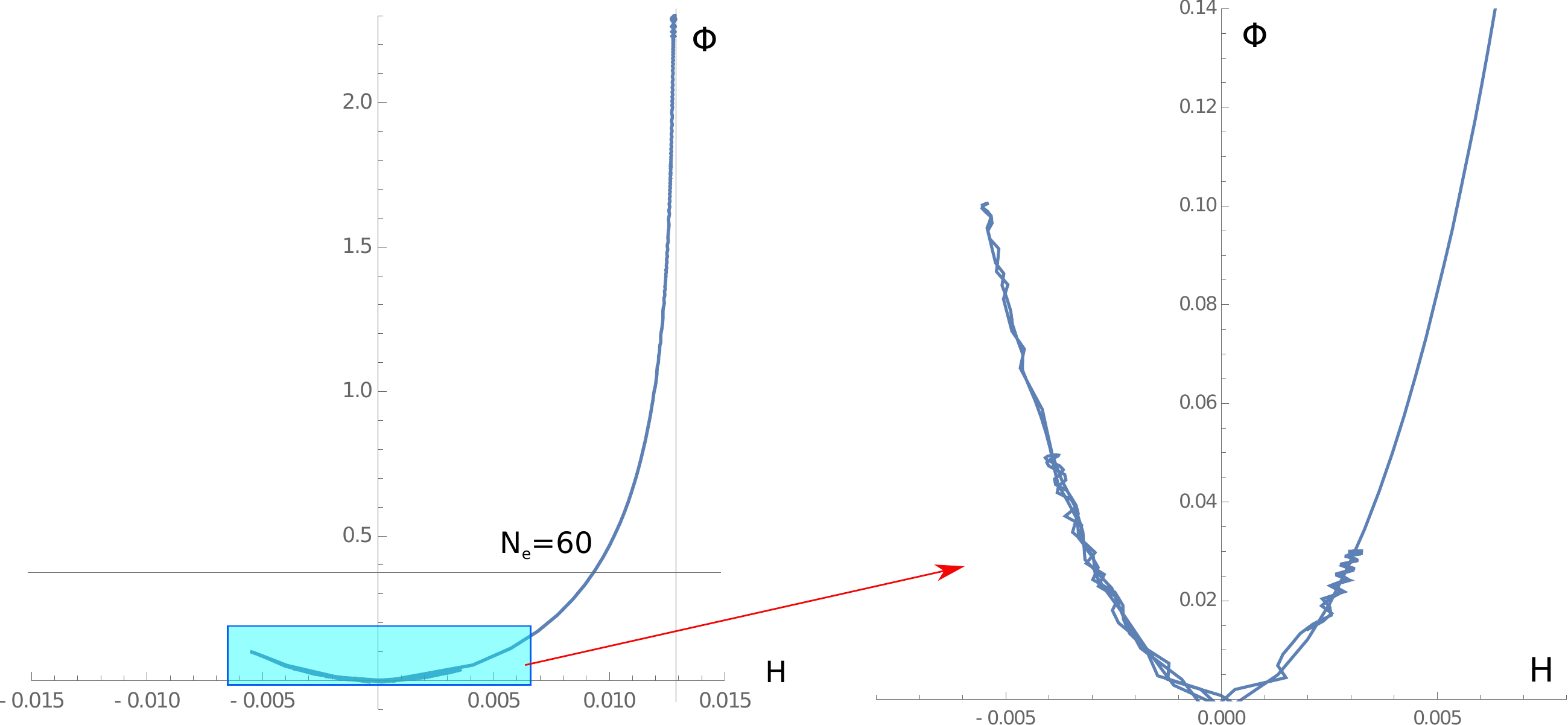}
\end{minipage}
\caption{Potential and inflaton trajectory in the $R^2$ model which
  provides a UV completion for the Higgs inflation. At the beginning
  of inflation, the rolling degree of freedom does not interact with
  the SM particles. At the end of inflation, the trajectory turns
  leading to an efficient reheating due to the production of the SM gauge
  bosons. Notice that in \cite{He:2018gyf} the potential looks
  somewhat different due to another choice of variables.
\label{fig:HR2-potential}
}
\end{figure}

{\bf 5.} The model predictions for cosmological parameters are similar to those of
the original Higgs-inflation\,\footnote{Or $R^2$-inflation\,\cite{starobinsky}, the
  difference in predictions of the two models is minuscule, for details
  see Refs.\,\cite{Bezrukov:2011gp,Gorbunov:2012ns}.} provided that
the single field approximation is valid for this model
\cite{Wang:2017fuy}. The inflationary trajectory lies in a deep valley
(see Fig.\,\ref{fig:HR2-potential})
which corresponds to the condition
\be
1-e^{-2\Phi}\cosh^2{H}-6\xi\sinh^2{H}=0,
\ee
so that only the first term in potential \eqref{action-5} contributes to
the energy density. This condition, however, places a constraint on
the two fields $\Phi$ and $H$. In general case,   
to obtain the amplitude of CMB fluctuations of order $10^{-5}$, one imposes a normalization condition \cite{Ema:2017rqn},
\be
\label{norm}
\beta+\frac{\xi^2}{\lambda}\simeq 2\times 10^{9}.  
\ee 
We can speak
about the Higgs-scalaron inflation as a UV completion of the Higgs
inflation if the CMB amplitude is actually defined by parameters of
the Higgs sector, $\lambda$ and $\xi$, rather than $\beta$. The heavy
degree of freedom indeed can be integrated out if
$\beta<\xi^2/\lambda~$\footnote{Notice that larger values of $\beta$
  (which correspond to the light scalaron) are not allowed by the
  normalization condition \eqref{norm} if $\xi$ is fixed. However, if $\xi$ is small the value of $\beta$ is fixed since it defines the amplitude of the scalar perturbations.}. In this case, the
predictions for the tilt of the scalar perturbation spectrum $n_s-1$
and tensor-to-scalar ratio $r$ are of
the standard form\footnote{This fact allows to distingiush this model from other UV completions for Higgs inflation suggested in the literature \cite{Giudice:2010ka,Lee:2013nv, Lee:2018esk}. These works consider an addition of the extra scalar field. The cosmological predictions in this case typically depend on the parameters of this hidden scalar.} \cite{Bezrukov:2007ep},
\be
\label{R2-Higgs-predictions}
n_s=1-\frac{2}{N_e},\qquad r= \frac{12}{N_e^2}
\ee
with $N_e=50\div 60$ being a number of e-foldings of inflation which
depends slightly on the reheating temperature. These predictions fall
right in the ballpark of the region allowed by the Planck experiment \cite{Ade:2015lrj}. No significant isocurvature and non-gaussianity is expected since the mass of the orthogonal direction is significantly larger than the Hubble scale (${\cal H}\sim 1.5\times 10^{13}$ GeV).
 
\begin{figure}[htb]
\begin{center}
\includegraphics[width=0.5\textwidth]{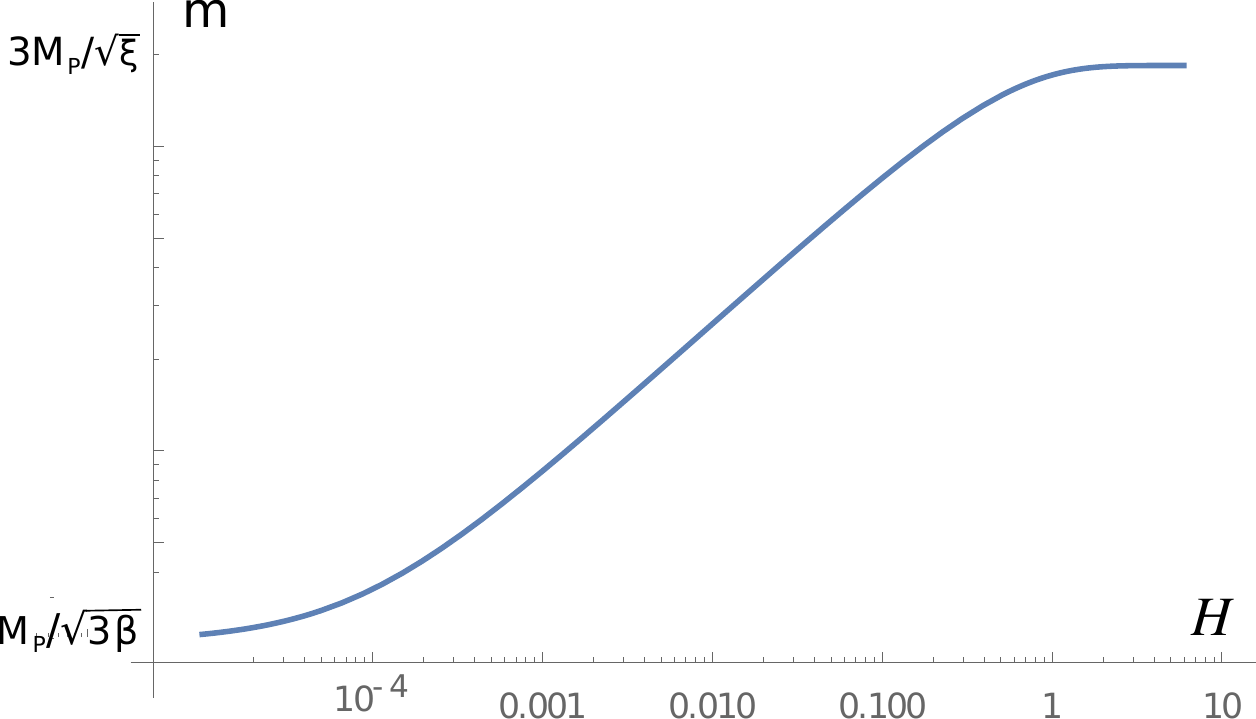}
\caption{The dependence of the effective mass of the isocurvature mode
  (orthogonal to the inflaton trajectory) on the value of field
  $H$. Notice that during inflation this mass is of order
  $M_P/\sqrt{\xi}$ while at smaller $H$ it becomes
  $M_P/\sqrt{3\beta}\lesssim M_P/\xi$. This behaviour is similar to
  the field-dependent cutoff scale in the Higgs inflation
  \cite{Bezrukov:2010jz}, something that is expected since this heavy
  degree of freedom provides a UV completion.
\label{fig:mass}
}
\end{center}
\end{figure}
 
\begin{figure}[h]
\includegraphics[width=0.9\textwidth]{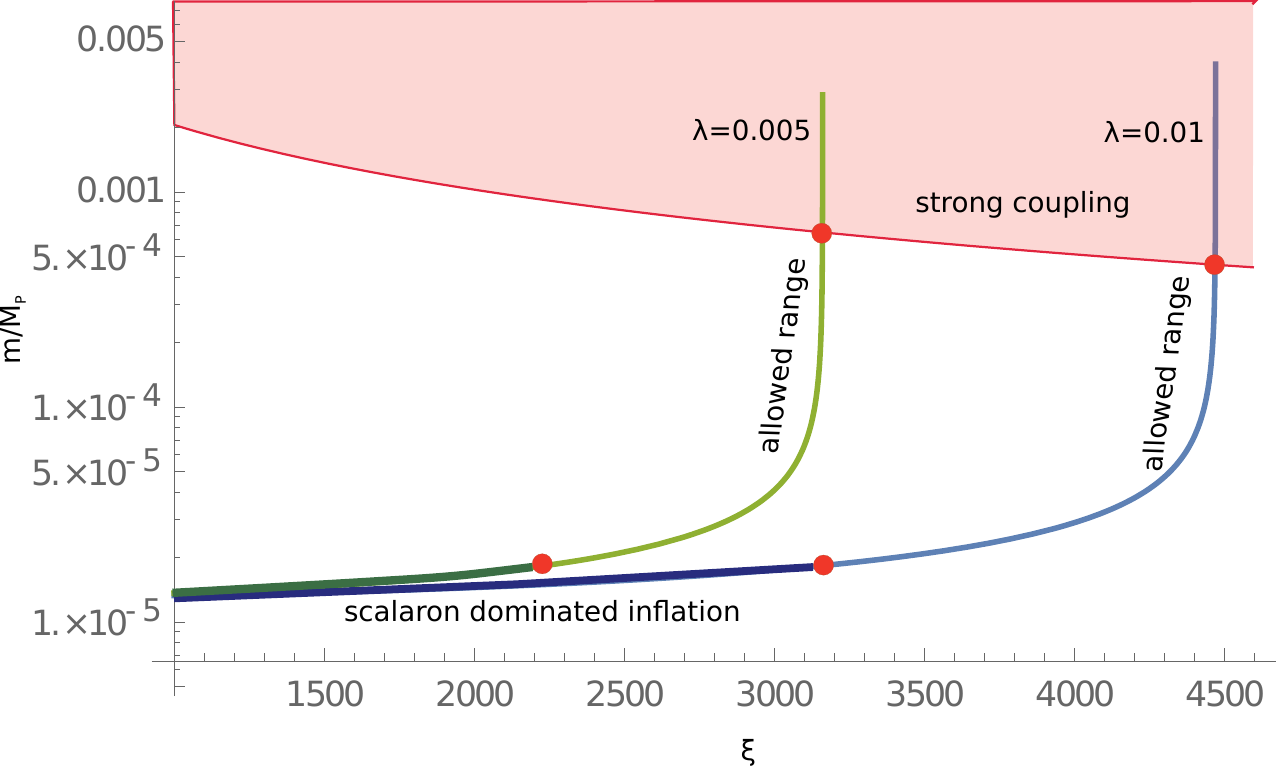}
\caption{The scalaron mass range for two reference values of the Higgs self-coupling during inflation. The allowed range is shown between the two (red) dots on the curve characterizing the dependence of the scalaron mass $m$ on the Higgs self-coupling $\xi$. This dependence comes as a result of the reqiurenment \eqref{norm}. The parts of the curves which lie below the lowest dots also correspond to the viable inflation model (scalaron dominated inflation) but this model cannot be considered as a UV completion of the original Higgs inflation.}
\label{fig:scalaron mass}

\end{figure}

Summarising the bound \eqref{norm} and perturbativity condition \eqref{perturbativity} on the parameter $\beta$ we can write,
\be
\label{bounds}
\frac{\xi^2}{4\pi}<\beta < \frac{\xi^2}{\lambda}.
\ee
Thus, with typical value of $\lambda\sim 0.01$ at large values of
the Higgs field 
the remaining window for parameter $\beta$ (which determines the scalaron
mass \eqref{mass}) is about three orders of magnitude. Consequently,
for the reference value $\lambda=10^{-2}$, the scalaron mass is in
the interval $5\times10^{13}~\text{GeV}<m<1.5\times 10^{15}$ GeV. The relevant part of the model parameter space is outlined in Fig. \ref{fig:scalaron mass}). 

{\bf 6.}
In this part of the Letter, we show that the $R^2$ term can cure the
strong coupling problem not only in the Higgs inflation. A possible
scale invariant extension of this model known as Higgs-dilaton
inflation \cite{GarciaBellido:2011de} also suffers from the similar
problem with the low cutoff scale. This model yields the Planck mass and naturally small Higgs mass from the spontaneous breaking of the scale symmetry. Under certain choice of parameters, it provides a viable inflationary stage. However, in this model, the Higgs field has to be coupled to gravity with large $\xi$ which again leads to the strong coupling scale about $M_P/\xi$, the same as in the original Higgs inflation.
 
The action of the Higgs-dilaton model completed with $R^2$ term reads,
\be
S=\int{d^4 x \sqrt{-g}\, \left[\frac{1}{2}[\beta R^2+(\d_{\mu} X)^2 - \xi X^2 R - \xi' h^2 R+(\d_{\mu} h)^2] - \frac{\lambda}{4}(h^2-\alpha^2 X^2)^2\right]}
\ee
In the Einstein frame this action can be written in such variables that the Higgs field direction becomes canonically normalized (see also Ref. \cite{Gorbunov:2013dqa} for a different choice of field variables where the scalaron field is canonical),
\be
\begin{split}
L=&\frac{1}{2}\l(\d H)^2+\cosh^2 {H} (\d \varphi)^2+\cosh^2{H}\cosh^2{\varphi}(\d\rho)^2\r-\\
-&\frac{1}{4}\l\lambda(\sinh^2{H}-\alpha^2\sinh^2{\varphi}\cosh^2{H})^2+\frac{1}{36 \beta}(1-6\xi\sinh^2{\varphi}\cosh^2{H}-6\xi'\sinh^2{H})^2\r.
\end{split}
\ee            
Here field $\rho$ plays a role of the Goldstone boson of the
broken scale invariance. It does not contribute to the potential. The
field $\varphi$ corresponds to the scalar degree of freedom coming from
gravity. The field $H$ is the only field coupled to the
gauge and fermion sectors of the SM with the interaction term
exactly of the form \eqref{GB}. Consequently, there is no strong-coupling issue in
the gauge sector of the model, as well as in the scalar-gravity sector. The cutoff scale of such model is again pushed up to the Planck scale.
 
\begin{figure}[h]
\includegraphics[width=0.9\textwidth]{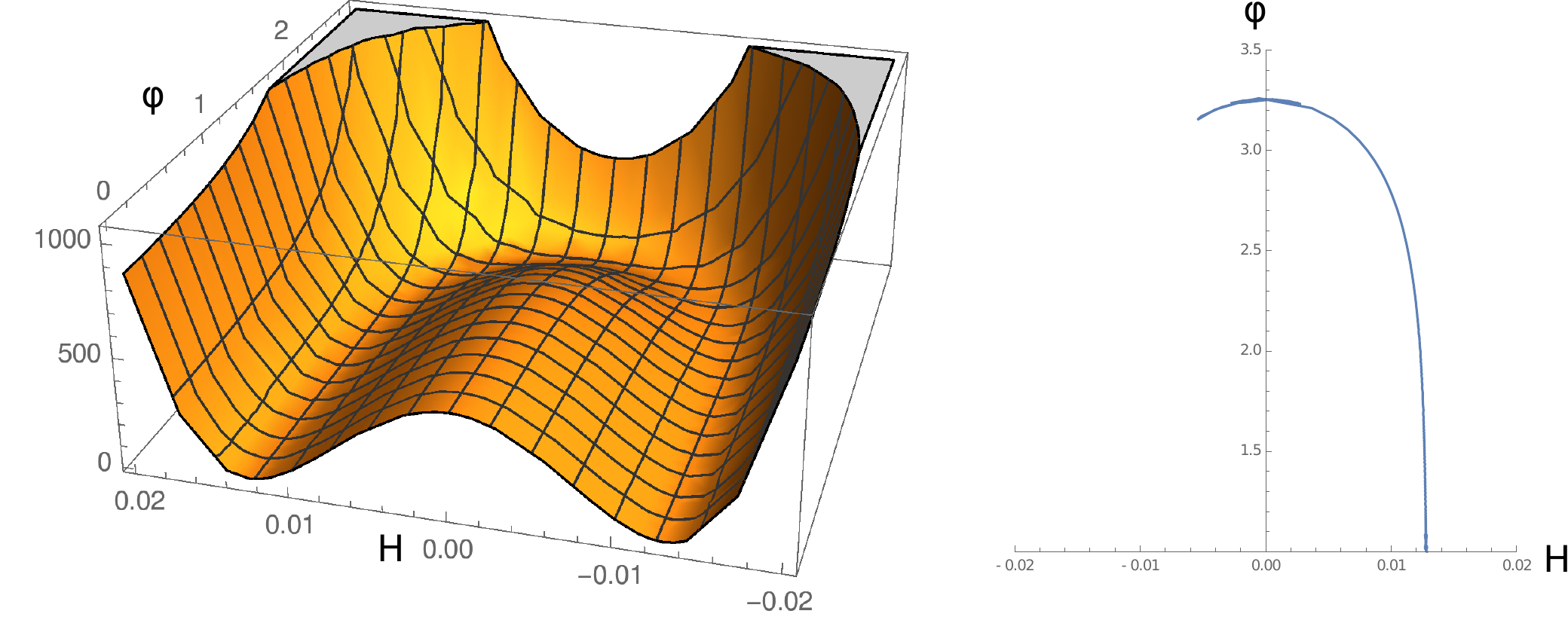}
\caption{Scalar potential (left panel) and inflaton trajectory (right panel) in Higgs-dilaton inflation
  with $R^2$ term.
\label{fig:HR2D-potential}
}
\end{figure}
 
The inflaton potential looks similar to the $R^2$-Higgs case. Again,
the inflationary stage can be effectively described as the single
field rolling inside the valley, under the conditions \eqref{bounds}
on $\beta$, see Fig.\,\ref{fig:HR2D-potential}. The predictions for
spectral parameters are the same as in the Higgs-dilaton model
\cite{GarciaBellido:2011de}: the scalar tilt depends on the value of $\xi$, 
\be
n_s=1-8\xi\coth{4\xi N_e}.
\ee
Therefore, in order to satisfy Planck limits \cite{Ade:2015lrj}, we need $\xi\lesssim 0.004$. The CMB amplitude can be obtained under the same condition as in \eqref{norm}.
 
\begin{figure}[!htb]
\begin{minipage}[h]{0.5\linewidth}
\includegraphics[width=\textwidth]{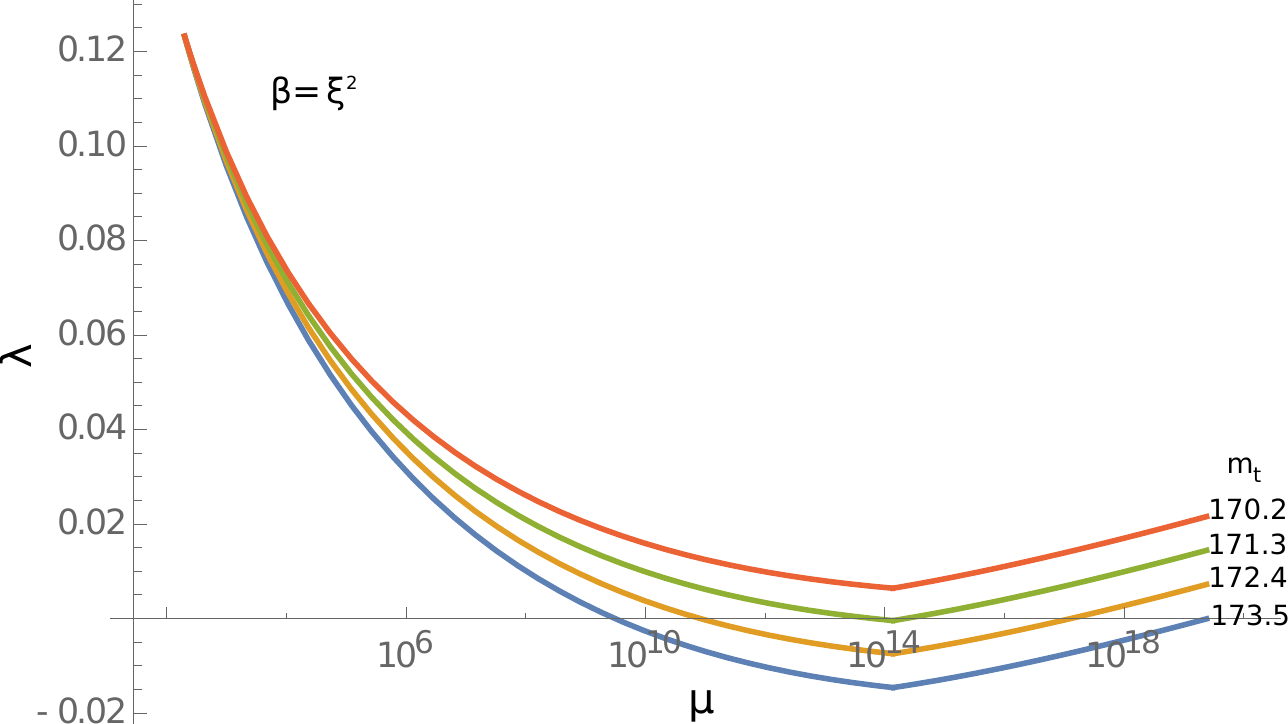}
\end{minipage}
\begin{minipage}[h]{0.5\linewidth}
\includegraphics[width=\textwidth]{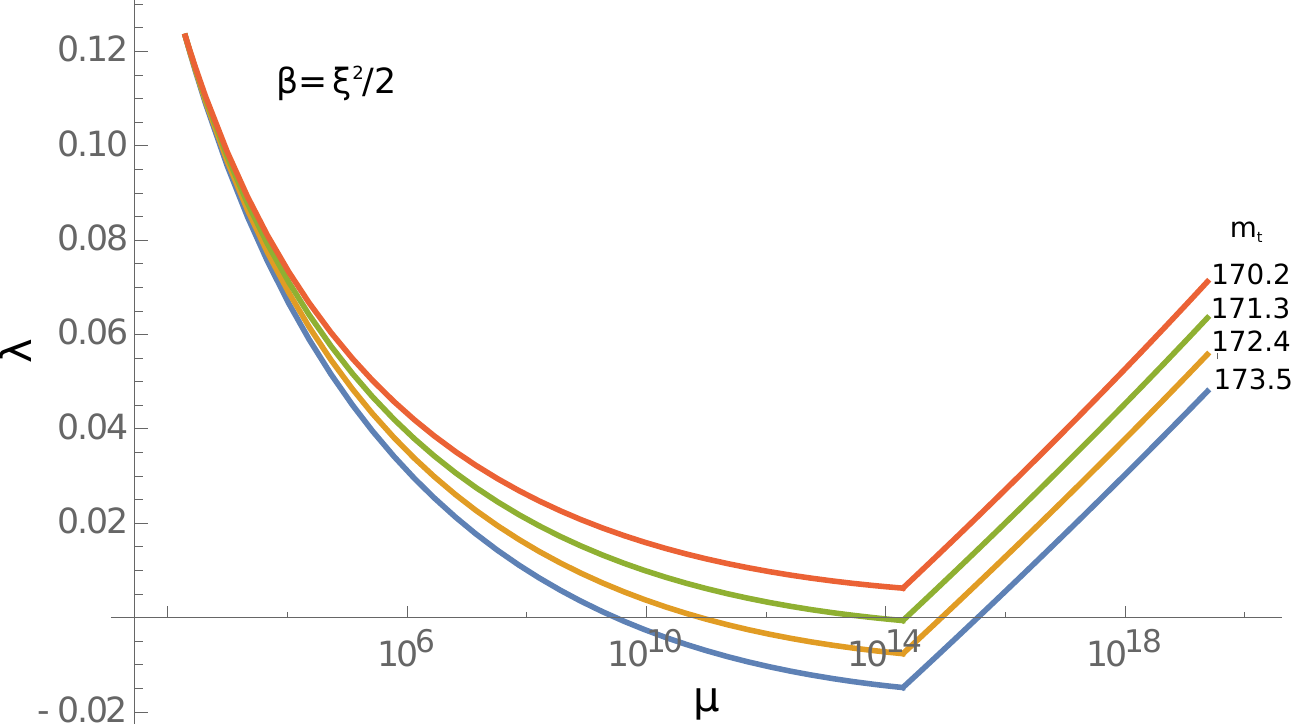}
\end{minipage}
\caption{The dependence of the Higgs self-coupling $\lambda$ on the
  renormalization scale $\mu$ with the scalaron one-loop impact
  included. The latter affects the running starting from the scale of
  order the scalaron mass \eqref{mass}. Here we plot the running of the parameter $\lambda$ which stands in the potential \eqref{action-4}. Due to the matching condition \cite{Ema:2017rqn}, this is exactly the paramer describing the low energy Higgs scattering.  
\label{fig:vac}
}
\end{figure}

{\bf 7.} Finally, as an extra bonus, the introduced $R^2$ term can
improve the stability of the Higgs potential. The latter is known to
take negative values at large fields if the central value of the top quark mass is
considered, for details see \cite{Bednyakov:2015sca}. The top Yukawa
coupling
contributes to the renormalization group running
of the Higgs self-coupling $\lambda$, such that it is hard to reach
the positive energy density during inflation (see \cite{Bezrukov:2017dyv}). The
scalaron provides a positive one-loop contribution to the
beta-function of $\lambda$ \cite{Avramidi:1986mj},
\be \delta
\beta_{\lambda}=\frac{1}{16\pi^2}\frac{2\xi^2(1+6\xi)^2}{9\beta^2}\,.
\ee 
Thus, in presence of the $R^2$-term the stability of the Higgs
potential can be secured for larger values of the top quark mass $m_t$
(see Fig.\,\ref{fig:vac}).
 
Notice also that if $\lambda$ is negative in some region of large
fields, the Higgs field would stay in the false vacuum during
reheating. However, due to the large reheating temperature, the
thermal corrections could finally bring the Higgs to the SM vacuum
\cite{Bezrukov:2014ipa}. While the detailed study of this process in
our model is required, we expect that the domain of the top quark
masses consisting with viable inflation becomes wider than in the minimal Higgs inflation.

{\bf 7.} To conclude, we show that a new gravitational scalar degree
of freedom can improve the models that suffer from the strong coupling
problem arising significantly below the Planck scale. With the
scalaron added, these models become theoretically self-consistent
cosmological models with inflation and reheating below the Planck
scale. We observe that under certain conditions on the scalaron mass,
introduction of this degree of freedom does not spoil the predictions
of the Higgs and Higgs-dilation inflation. Moreover, the model allows for 
a {\it consistent} description of the particle production after
inflation. We leave the detailed study of the reheating in these
models for future work.
 
\vskip 0.2cm

The authors are grateful to F. Bezrukov, M. Shaposhnikov, S. Sibiryakov for valuable discussions. This work was supported by Russian Science Foundation grant 14-22-00161.
 

\end{document}